\documentstyle[pra,aps,rotate,epsf]{revtex}

\begin{document}

\newcommand{\rr}{{\bf r}}
\newcommand{\la}{\langle}
\newcommand{\ra}{\rangle}
\newcommand{\Om}{\Omega}
\newcommand{\be}{\begin{equation}}
\newcommand{\ee}{\end{equation}}
\newcommand{\ba}{\begin{eqnarray}}
\newcommand{\ea}{\end{eqnarray}}
\newcommand{\kap}{{\mbox{\boldmath $\kappa$}}}
\newcommand{\subkap}{{\mbox{\scriptsize\boldmath $\kappa$}}}
\newcommand{\half}{\frac{1}{2}}
\newcommand{\quarter}{\frac{1}{4}}
\newcommand{\eighth}{\frac{1}{8}}
\newcommand{\R}{{\bf R}}
\newcommand{\Q}{{\bf Q}}
\newcommand{\x}{{\bf x}}
\newcommand{\qr}{{\Q \cdot \R}}
\newcommand{\xiR}{\xi(\R)}
\newcommand{\etaR}{\eta(\R)}
\newcommand{\DxiR}{\Delta\xiR}
\newcommand{\DZR}{\Delta Z(\R)}
\newcommand{\p}{{\rm Pr}}

\title{Atom Scattering from Disordered Surfaces in the Sudden
Approximation: Double Collisions Effects and Quantum Liquids
}

\author{Daniel A. Lidar (Hamburger)}
\address{
Chemistry Department\\
University of California\\
Berkeley, CA 94720\\
USA  
}

\maketitle

\begin{abstract}
The Sudden Approximation (SA) for scattering of atoms from surfaces is
generalized to allow for double collision events and scattering from
time-dependent quantum liquid surfaces. The resulting new schemes retain the
simplicity of the original SA, while requiring little extra
computational effort. The results suggest that inert atom
(and in particular He) scattering can be used profitably to study
hitherto unexplored forms of complex surface disorder.\\
\end{abstract}


\markboth{{Daniel A. Lidar}}{{{\it Surf. Sci.} 1998}}

\section{Introduction}

Structurally disordered surfaces have been a subject of great interest
for some time now. Of special interest are epitaxially grown
films, liquid surfaces, and amorphous surfaces.  In
epitaxial growth for example, metal or semiconductor atoms are adsorbed on a
surface under thermal conditions, to form two- and
three-dimensional structures on top of it. The physical and chemical
properties are determined by the final form of these structures. These
may be of dramatic importance, e.g, in the production of electronic
devices. One of the most exciting aspects of epitaxial growth
kinetics, is that it {\em {creates disordered structures}} in the
intermediate stages. The disorder manifests itself in the formation of
various types of clusters or diffusion limited aggregates on top of
the surface. These structures may be monolayers (usually at high
temperatures, when the diffusivity is large, or at coverages
significantly below one monolayer), in which case the disorder is
two-dimensional, or they may be composed of several layers, giving
rise to disorder in three dimensions. Epitaxially grown structures of
this type 
offer an exceptional opportunity for both experimental and theoretical
studies of disorder. No satisfactory and comprehensive theory of the
epitaxial growth process is as of yet available, much due to the
absence of reliable interaction potentials for the system. The situation with respect to liquid and amorphous
surfaces is similar: very little is known at this point about their structure. Progress at
this stage thus hinges critically on data available from
experiments. An important experimental technique is thermal atom
scattering, and in particular He
scattering \cite{Engel:82,Comsa:book,Benny:review2,Ibach,Hulpke:92,Benny:review1,Rieder:94,me:CAMP}.
The main advantage offered by He scattering is complete {\em surface}
sensitivity, as thermal He atoms do not penetrate into the bulk, unlike other
scattering techniques such as neutron or X-ray scattering, or
LEED. Another important advantage is that He scattering is highly
non-intrusive, due to the inertness and low energy of the He atoms. The
latter also means that He scattering is really a {\em diffraction}
experiment at the typical meV energy scale at which most experiments
are performed, with sensitivity to {\em atomic-scale} features.
The interpretation of He scattering experiments is, however, rather
involved due to the complicated interaction between the He atom and
the surface.

A highly successful theoretical method in the study of He
scattering is the Sudden Approximation (SA), introduced in the context of
atom-surface scattering by Gerber, Yinnon and Murrell in 1978
\cite{Benny:Sud1}. It is to date one of the most useful tools in the
field, and has been reviewed by Gerber \cite{Benny:review2,Benny:review1}. While the original
formulation of the SA applied to periodic surfaces, it was extended by Gersten {\it et al.} to deal with scattering
from disordered surfaces \cite{Benny:disorder}. The SA was subsequently
used to study a large variety of disordered systems, such as isolated
adsorbates on crystalline surfaces
\cite{Benny:RBs,Benny:Sud2,Peppino,me:optical}, mixed overlayers of
Xe+Ar \cite{Yinnon:84,Benny:Sud2} and Xe+Kr on Pt(111) \cite{Benny:XeKr},
randomly corrugated hard walls \cite{Benny:Sud3}, vacancies and CO
adsorbates on Pt(111) \cite{Benny:3,Benny:4}, percolation lattices of
substitutionally disordered Xe+Kr monolayers \cite{Gerber:91}, amorphous mixed
monolayers and liquids \cite{Benny:liquids}, compact islands and
diffusion limited
aggregates \cite{me:Heptamers,me:Ag-systems,me:SSL97}, and
fractals \cite{me:fractals,me:PRE96}. Common to many of these studies
was the achievement of detailed insight into the collision dynamics,
the prediction and understanding of interesting features in the
angular intensity distribution of the scattered atoms. In several
cases it was shown that the angular intensity distribution exhibits
nonspecular maxima of two types: Several of the peaks are
defect-induced rainbows effects, while others (at angles nearer to the
specular) are Fraunhofer diffraction interferences.  Both types of
peaks contain useful, largely complementary, information on defect
geometry and on the He/defect interaction.
Several further extensions of the SA include the study of Schinke and
Gerber which treated phonons
and the resulting Debye-Waller attenuation \cite{Benny:DW}; and Hinch's
work \cite{Hinch:89} which extended the SA to encompass larger
parallel momentum 
transfers.

The main purpose of this work is to extend the SA in two useful directions:
the inclusion of double collision processes, and scattering from
time-dependent quantum liquid surfaces. Unfortunately, at the time of
writing experiments are unavailable for comparison with the results
obtained here. The developments will therefore be primarily
methodological, in anticipation of experimental data. It is hoped that
the results obtained here will motivate He scattering experiments on
disordered solid and liquid surfaces. As demonstrated by this work and
others before it, He
scattering can provide a wealth of information on disordered surface
structure and dynamics.

For reasons detailed in
Secs.~\ref{SA} and \ref{IS}, the SA in its original formulation cannot
include double collision events. These, however, are important in many
cases, especially when the surface corrugation is large. An analysis
of ``rainbow'' peaks arising due to double 
collisions and performed here for the first time, shows that these are
distinct from single-collision 
rainbows in that they have an incidence-energy dependence. This should
be clearly observable experimentally, e.g., in scattering from CO
defects on Pt(111). Multiple scattering by defects has been considered
in the past, e.g., by J\'{o}nsson, Weare and Levi \cite{Jonsson2} who
derived interesting results concerning the dependence of the specular
intensity on coverage for different models of disorder (applied to Xe and CO); by Kara and Armand \cite{Kara:85}, who
considered elastic scattering 
from a Cu(110) surface covered with randomly distributed copper
atoms; A similar multiple scattering theory was presented by Armand and Salanon
\cite{Armand:89:2}, who showed that for randomly distributed
defects, the incoherent cross section is
equal to the product of a form factor and a Fourier transform of the
two body correlation function of the defects' positions. Common to these works is a multiple scattering expansion of
the Lipmann-Schwinger equation for the T-matrix, in powers of the
number of He scattering events \cite{Jonsson2}, a perturbation potential
due to defects \cite{Kara:85}, or the number of defects involved in
the scattering \cite{Armand:89:2}. The approach to be presented here
is somewhat similar to that of \cite{Jonsson2}, in that the He atom is
assumed to undergo a second collision. However, the combination of the
SA and the Born approximation used here is new, and leads to
expressions of significantly greater computational simplicity than
that of the general multiple scattering theory.

The study of liquid surfaces is an exciting and novel topic in
its own right \cite{King:93,Ronk:96,Nathanson:96}, and to date there
are very few theoretical methods available for this purpose, in which
the surface degrees of freedom are treated classically
\cite{me:CAMP,Benny:unpublished}. The extension of the SA 
that is presented here, in Sec.\ref{TDSCF-SA}, treats the surfaces as
a quantum liquid, within a time-dependent self-consistent-field
(TDSCF) approach. This methodological part of the paper serves to
generalize the SA 
to include time-dependent processes and thus renders this useful
method significantly more widely applicable. Concluding remarks are
brought in Sec.\ref{conclusions}.

\section{Brief Review of the Original Sudden Approximation}
\label{SA}

Consider a He atom with mass $\mu$ incident upon a surface with wavevector ${\bf k}
\!=\! ({\bf K},k_z)$. $\hbar {\bf K} \!=\! \hbar(k_x,k_y)$ and $\hbar {\bf K}'$ are respectively the intial and
final momentum components parallel to the surface, and $\hbar k_z$ is
the incident momentum normal to the surface. The position of the He
atom is ${\bf r} \!=\! ({\bf R},z)$, where ${\bf R} \!=\!
(x,y)$ is the lateral position. The SA is valid when the collisional momentum
transfer ${\bf q} \!=\! {\bf K}'-{\bf K}$ in the direction
parallel to the surface is much smaller than the momentum transfer
normal to the surface: $2 k_z \gg |{\bf q}|$. This condition is
satisfied close to specular scattering and energy $E \!=\! (\hbar {\bf
k})^2/(2m)$, and
moderate surface corrugations. When it holds, one can approximately
consider the scattering along $z$ as occurring at fixed ${\bf R}$. Then
if $\psi$ is the He wavefunction, it satisfies a 
Schr\"{o}dinger equation where the dependence on ${\bf R}$ is {\em
adiabatic}:

\be
\left[ -{\hbar^2 \over 2\mu} {d^2 \over dz^2} + V_{\bf R}(z) \right]
\psi_{\bf R}(z) = E\psi_{\bf R}(z) .
\label{eq:SE}
\ee

\noindent Here $V_{\bf R}(z)$ is the He-surface interaction
potential and no inelastic channels are included. This means that each
surface point ${\bf R}$ gives rise to an elastic real phase shift
$\eta({\bf R})$, which can be evaluated in the WKB approximation from
Eq.(\ref{eq:SE}) as:

\ba
\eta({\bf R}) = \int_{\xi({\bf R})}^{\infty} dz\:
\left[\left( {k_z}^2 - {2m \over \hbar^2}V_{\bf R}(z) \right)^{1/2} -k_z
\right] - k_z\,\xi({\bf R}) ,
\label{eq:eta}
\ea

\noindent where $\xi({\bf R})$ is the classical turning point
pertaining to the integrand in Eq.(\ref{eq:eta}). The phase shift in
turn yields the S-matrix as: ${\cal S}({\bf R}) \!=\! \exp[2i \eta({\bf
R})]$. The ${\bf R}$ coordinate is conserved in this picture so the 
S-matrix is diagonal in the coordinate representation:

\[
\langle {\bf R}'| {\cal S} | {\bf R} \rangle = e^{2i \eta({\bf R})}
\delta({\bf R}'-{\bf R}) .
\]

\noindent Experimentally one measures probabilities $|{\cal
S}({\bf K} \rightarrow {\bf K}')|^2$ for ${\bf K}
\rightarrow {\bf K}'$ transitions. To obtain these $\langle
{\bf R} | {\bf K} \rangle \!=\! \exp(i {\bf K} \!\cdot\! {\bf
R})/\sqrt{A}$ (where $A$ is the area of the surface) can be used, to find:

\ba
\langle {\bf K}'| {\cal S} | {\bf K} \rangle =
\int d{\bf R}' \: d{\bf R}
\langle {\bf K}'|{\bf R}' \rangle
\langle {\bf R}'| {\cal S} | {\bf R} \rangle 
\langle {\bf R} |{\bf K}  \rangle =
{1 \over A} \int d{\bf R} e^{-i {\bf q} \cdot {\bf R}}
e^{2i \eta({\bf R})} .
\label{eq:Sud}
\ea

\noindent This is the well-known expression for the SA scattering
amplitude \cite{Benny:Sud1,Benny:disorder}.

\section{The Iterated Sudden Approximation for Double Collision Events}
\label{IS}

\subsection{Background}

\begin{figure}
\hspace{6cm}
\epsfysize=6cm
\epsfxsize=14cm
\epsffile{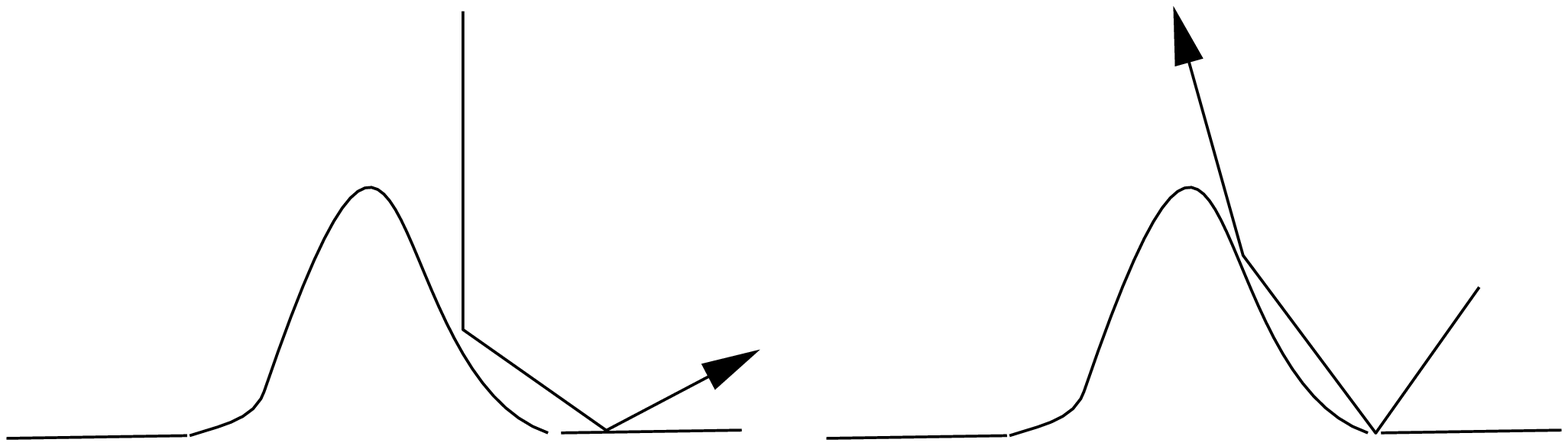}
\caption{
On the left, the first collision strikes the adsorbate, the second
strikes the surface. The reverse sequence occurs on the right.}
\label{fig:collisions}
\end{figure}

The SA developed in Sec.\ref{SA} can inherently describe
only {\em single collision} events between the incident particle and
the surface. This is because the SA assumes that the
momentum transfer in 
the direction parallel to the surface is much smaller than in the direction
perpendicular to it, i.e., the surface is assumed ``not too corrugated''. A
more formal way to see this is by realizing that the SA essentially treats the
${\bf R} = (x,y)$ coordinate as {\em adiabatic}, which implies that each
trajectory takes place at constant ${\bf R}$. Clearly, no double collisions
can occur under such conditions. However, double collisions may take
place, e.g., when an incident atom is scattered off a defect onto the
surface, or in the opposite order, as depicted in
Fig. \ref{fig:collisions}. The purpose of this section is to
develop an ``Iterated Sudden'' (IS)
approach, which can be interpreted to include double collision effects as
well.

Starting from the first principles of non-relativistic scattering
theory, the exact expression for the T-matrix element is \cite{Taylor}:

\begin{equation}
t({\bf k} \rightarrow {\bf k}') = \langle {\bf k}' | V | {\bf k}+
\rangle ,
\label{eq:Tmatrix}
\end{equation}

\noindent where $|{\bf k}+ \rangle = \Omega_+ \, | {\bf k}
\rangle$. Here $|{\bf k} \rangle$ is the incident state
($t\!=\!-\infty$) and $\Omega_+$
is the M{\o}ller operator, which propagates $|{\bf k} \rangle$ to the
state at the time of collision at $t\!=\!0$. In the Born
approximation it is assumed that $| {\bf k}+ \rangle \approx | {\bf k}
\rangle$ \cite{Taylor}, which yields $t({\bf k} \!\rightarrow\!
{\bf k}') \!=\! \langle {\bf k}' | V | {\bf k} \rangle$. The IS, on
the other hand, consists in setting:

\begin{equation}
| {\bf k}+ \rangle \approx | {\bf k}^{\rm sud} \rangle \:\: ;
\:\:\:\:\:
| {\bf k}^{\rm sud} \rangle = S^{\rm sud} \, | {\bf k} \rangle ,
\label{eq:IS1}
\end{equation}

\noindent and where $S^{\rm sud}$ is the S-matrix operator in the SA
[Eq.(\ref{eq:Sud})]. Hence: 

\begin{equation}
t({\bf k} \rightarrow {\bf k}') \approx t^{\rm sud}({\bf k}
\rightarrow {\bf k}') = \langle {\bf k}' | V | {\bf k}^{\rm sud}
\rangle 
\label{eq:Tmatrix-sud}
\end{equation}

Eq.(\ref{eq:Tmatrix-sud}) can be argued to describe a double
collision process as follows. In the exact expression
Eq.(\ref{eq:Tmatrix}), $| {\bf k}+ \rangle$ represents the actual
state at $t=0$ (``the moment of collision''), as it evolved from
$t=-\infty$. However, $| {\bf k}^{\rm sud}
\rangle$ is the state at $t=+\infty$,\footnote{The M{\o}ller operator
$\Omega +$ propagates the in-state $|\psi_{\rm in} \rangle$ to the
actual state at $t=0$. The $S$ operator propagates $|\psi_{\rm in}
\rangle$ to the actual state at $t=+\infty$ \cite{Taylor}.} in
the SA. Thus Eq.(\ref{eq:IS1}) implies setting:\\ {\it ``actual state
at collision'' = ``approximate state much after single collision''}.\\
In other words, at the moment of collision, the wave function is taken
to have already undergone a prior collision.

A qualitative argument can be given to justify the validity of this
approximation. The Lipmann-Schwinger equation for the exact wave function
$\psi$ is:

\begin{equation}
\psi = \Phi_{\rm in} + G_0 \, V \, \psi .
\label{eq:LS}
\end{equation}

\noindent Now, an equivalent way to view the IS is by setting $\psi
\approx \psi^{\rm sud}$ in the right-hand side of Eq.(\ref{eq:LS}): 

\[
\psi = 
 \left[ \Phi_{\rm in} + G_0 \, V \, \psi^{\rm sud} \right]  \, + \, G_0 \, V
 \, \Delta \psi ,
\]

\noindent where $\Delta \psi \equiv \psi - \psi^{\rm sud}$. Thus by
 assuming $\psi \approx \psi^{\rm sud}$, $G_0 \, V \, \Delta \psi$ is
 neglected. To estimate the magnitude of this term, consider two
 length-scales: First, it is clear that $G_0 \, V \,
\Delta \psi$ is small at {\em large} distances from the surface, since $V$
becomes small. Secondly, $G_0 \, V \, \Delta \psi$ is also small {\em
close} to the surface, due to the nature of the SA: the SA
diagonalizes the Schr{\"{o}}dinger equation $(T+V)\psi = E \, \psi$,
$E_{ij} = \epsilon_{ii} \,
\delta_{ij}$ by taking the energy matrix $E$ to be a {\em scalar} matrix: $E =
\epsilon_0 \, I$, $\epsilon_0 = \max_{i} \epsilon_{ii}$ \cite{Benny:Sud1}. Hence, the SA essentially assumes 
that the other ($i>0$) diagonal elements $\epsilon_i$ are negligible
in comparison to the elements of the diagonalized $V$. This condition
is well satisfied {\em close} to the surface, where $V$ is
largest. Hence one expects $\Delta \psi \approx 0$ to good accuracy
for both large and small distances from the surface.  Interpolating, one can
reasonably expect $G_0 \, V \, \Delta \psi$ to be negligible at {\em all}
distances.

\subsection{Calculation of the T-Matrix Elements}
The calculations in this section will be performed in two ways: the
first will lead to a result which can be used in numerical
simulations, the second will employ some simplifications in order to
cast the results in a form amenable to analytic analysis.

\subsubsection{Wave Function Iterated Sudden}
From the close-coupling expansion it follows \cite{Benny:disorder}
that the Sudden wave function in the continuum case is:

\[
|\psi_{\rm in}^j ({\bf r}) \rangle = \int d{\bf q} \: |\psi_{\bf q}^j ({\bf r})
\rangle ,
\]

\noindent where $j=1,2$ depending on whether one is in the classically
allowed or forbidden region, and:

\[
\langle {\bf r}|\psi_{\bf q}^j ({\bf r}) \rangle = \Phi_{\bf q}^j (z) \,
e^{i({\bf K}_{\rm in}+{\bf q}) \cdot {\bf R}}
\]

\[
\Phi_{\bf q}^j (z) = \alpha \int d{\bf R}\: e^{-i \, {\bf q} \cdot {\bf
R}} \, \Phi_{\bf R}^j (z)
\]

\noindent The last equation is where the Sudden approximation is made:
$\psi_{\rm in}^j ({\bf r})$ was replaced by $\Phi_{\bf R}^j (z)$, i.e,
the ${\bf R}$ coordinate is treated as an adiabatic parameter. This is
now further evaluated in the WKB approximation:

\[
\Phi_{\bf R}^j (z) \approx {1 \over {\sqrt {k ({\bf r})}}} \exp\left[ -i
\left( \int_{\xi({\bf R})}^z dz' \: k ({\bf R},z') \, + \, \phi^j \right)
\right] .
\]

\noindent Here:

\[
k({\bf r}) = \sqrt{{k_{\rm in}^z}^2 - {{2\, m} \over {\hbar^2}}
V({\bf r})}.
\]

\noindent In the classically allowed (forbidden) region $\phi^{(1)}
\!=\! -\pi /4$ and $k({\bf r})$ is real ($\phi^{(2)} = 0$; $k({\bf
r})$ is pure imaginary). The normalization ($\langle \psi_{\rm in}^j |
\psi_{\rm in}^j \rangle = 1$) is found after some algebra:

\[
| \alpha | = {1 \over {(2 \pi)}^2} \:  \left[ \int d{\bf r} \: |
\Phi_{\bf R}^j 
(z) |^2  \right] ^{-{1 \over 2}} \approx {1 \over {(2 \pi)}^2} \:
\left[  \int d{\bf r} \: {1 
\over {k({\bf r})}} \right] ^{-{1 \over 2}} .
\]

\noindent Now it is possible to evaluate the matrix element within the IS:

\begin{eqnarray}
\lefteqn{t^j ({\bf k}_{\rm in} \rightarrow {\bf k}_{\rm out}) = \langle {\bf k}_{\rm out} | V({\bf r}) | \psi_{\rm in}^j \rangle =} \nonumber \\
& & \int d{\bf q} \int d{\bf r} \, d{\bf r}' \: \langle {\bf k}_{\rm
out} | {\bf r} 
\rangle \, \langle {\bf r} | V({\bf r}) | {\bf r}' \rangle \, \langle {\bf r}'
| \psi_{\bf q}^j ({\bf r}) \rangle = \nonumber \\
& & \int d{\bf q} \int d{\bf r} \, d{\bf r'} \: {(2 \pi)}^{-{3 \over 2}} \,
e^{-i \, {\bf k}_{\rm out} \cdot {\bf r}} \, [V({\bf r}') \,
\delta({\bf r}-{\bf 
r}')] \, \Phi_{\bf q}^j (z') \, e^{i({\bf K}_{\rm in}+{\bf q}) \cdot
{\bf R}'} = 
\nonumber \\
& & {{\alpha} \over {{(2 \pi)}^{3 \over 2}}} \: \int d{\bf r}  \: V({\bf r})
\, e^{-i(\Delta {\bf K} \cdot {\bf R} \, + \, k_{\rm out}^z \,z)} \;
\int d{\bf R'} 
\, \Phi_{R'}^j (z) \; \int d{\bf q}\: e^{i\, {\bf q} \cdot ({\bf R}-{\bf R'})}
= \nonumber \\
& & \sqrt{2 \pi} \, \alpha \: \int d{\bf r} \: V({\bf r}) \, e^{-i(\Delta
{\bf K} \cdot {\bf R} \, + \, k_{\rm out}^z \,z)} \, \Phi_{\bf R}^j (z)
. 
\label{eq:general-t}
\end{eqnarray}

\noindent where $\Delta {\bf K} \equiv {\bf K}_{\rm out}-{\bf K}_{\rm
in}$. Employing
the WKB approximation for the wave-function $\Phi_{\bf R}^z (z)$, one has
finally for the transition amplitude from the disordered surface
within the IS:

\begin{eqnarray}
\lefteqn{t^j ({\bf k}_{\rm in} \rightarrow {\bf k}_{\rm out}) = } \nonumber \\
&&
\alpha \, e^{i \, \phi^j} \, \sqrt{2 \pi} \int d{\bf r} {1 \over
{\sqrt{k ({\bf r})}}} \, V({\bf r}) \,
\exp\left[ -i \left( \int_{\xi({\bf R})}^z dz' \: k ({\bf R},z') \,
+ \, \Delta {\bf K} \cdot {\bf R} \, + \, k_{\rm out}^z \,z \right)
\right] .
\label{eq:WF-t}
\end{eqnarray}

\noindent Comparing the IS expression
(\ref{eq:WF-t}) to the SA expression (\ref{eq:Sud}), note that the
computational complexity is essentially increased merely by an additional
$z$-integration. Considering the computational simplicity of the SA, the IS
offers an attractive improvement. A further discussion of the physical
significance of the IS result is given in Appendix \ref{app1}.

\subsubsection{Phase-Shift Iterated Sudden}
Equation (\ref{eq:WF-t}) is suitable for numerical applications. In
order to make further analytic progress, I now develop an alternative,
more approximate expression. This will involve calculating the
T-matrix element for the IS again, but assuming that {\em off}-shell (i.e., non energy conserving: ${|{\bf q}|}^2 \neq
{|{\bf q'}|}^2$) S-matrix elements are well
approximated by {\em on}-shell ones. The on-shell elements to be used are
just the SA elements from Eq.(\ref{eq:Sud}). That these are indeed on-shell,
follows simply from the fact that Eq.(\ref{eq:Sud}) represents the SA result
for {\em elastic} scattering. Starting again from
Eq.(\ref{eq:Tmatrix-sud}) [standard intermediate steps are omitted
from now on for brevity; the details are as in Eq.(\ref{eq:general-t})]:

\begin{eqnarray}
\lefteqn{t^j ({\bf k}_{\rm in} \rightarrow {\bf k}_{\rm out}) =
\langle {\bf k}_{\rm out} 
| V({\bf r}) \, S^{\rm sud} | {\bf k}_{\rm in} \rangle =} \nonumber \\
&&
{1 \over {(2 \pi)}^5} \, \int d{\bf q} \, d{\bf q}' \: \int d{\bf R}' \:
e^{i \, ({\bf K}_{\rm in} \, - \, {\bf q}) \cdot {\bf R}'} \, \int d{\bf r} \:
e^{-i \, ({\bf k}_{\rm out} \cdot {\bf r} \, - \, k_{\rm in}^z \, z \,
- \, {\bf R} \cdot {\bf q}')} S_{{\bf q}'{\bf q}} \, V({\bf r}) \nonumber
\end{eqnarray}

\noindent Due to the double integration over the unrestricted
intermediate momentum values ${\bf q}$, ${\bf q}'$, the S-matrix
element is off-shell. Approximating $S_{{\bf q}'{\bf q}}$ by the
on-shell expression Eq.(\ref{eq:Sud}) yields:

\ba
\lefteqn{t^j ({\bf k}_{\rm in} \rightarrow {\bf k}_{\rm out}) \approx}
\nonumber \\
&&
{1 \over {{(2 \pi)}^3} A} \int d{\bf q} \: \left( \int d{\bf R}'
\: e^{i \, ({\bf K}_{\rm in} \, - \, {\bf q}) \cdot {\bf R}'} \, e^{2i
\eta({\bf 
R}')} \right) \int d{\bf r} \: V({\bf r}) \, e^{i\,  \left[ {\bf R} \cdot
{\bf q} \, + \, k_{\rm in}^z \, z \, - \, {\bf k}_{\rm out} \cdot {\bf
r} \right] } .
\label{eq:on-shell}
\ea

\noindent And finally: 

\ba
t^j ({\bf k}_{\rm in} \rightarrow {\bf k}_{\rm out}) \approx
{1 \over {(2 \pi) \, A}} \: \int d{\bf r} \: V({\bf r}) \, e^{i
\,  \left[ 2\, \eta({\bf R}) \, - \, \Delta {\bf k} \cdot {\bf r}
\right] } .
\label{eq:PS-t}
\ea

\noindent The last result is the phase-shift equivalent of
Eq.(\ref{eq:WF-t}), in the 
{\em {on-shell approximation}}. This form lends itself more easily to
analysis, as will be demonstrated in Sec.\ref{examples}. Appendix
\ref{app2} provides a further discussion of the physical
significance of this result.

\subsection{Analysis}
\label{examples}
In this section I will analyze the expression for the scattering amplitude,
including double collisions [Eq.(\ref{eq:PS-t})], by considering a
simple model potential, of the form:

\[
V=V[z-\xi ({\bf R})] .
\]

\noindent
This is a rather general potential, in that the surface shape function
$\xi ({\bf R})$ can be chosen arbitrarily. More examples can be found
in Ref. \cite{me:thesis}.
The significance of the $V[z-\xi ({\bf R})]$ potential is
the following: $z=\xi 
({\bf R})$ is a two-dimensional surface in real space, and is clearly
also an equipotential surface. Thus $z = z_0 + \xi ({\bf R})$ defines
a continuum of equipotentials as a function of the height $z_0$ above
the surface plane. Hence $V[z-\xi ({\bf R})]$ can be interpreted as
the potential of a corrugated surface, with $\xi ({\bf R})$
representing the corrugation. Now, inserting $V$ into
Eq.(\ref{eq:PS-t}) one obtains:

\[
t({\bf k}_{\rm in} \rightarrow {\bf k}_{\rm out}) \propto \int d{\bf
r} \: e^{i \, (2 \eta ({\bf R}) \, - \, \Delta {\bf k} 
\cdot {\bf r})} \, V[z-\xi ({\bf R})] .
\]

\noindent Transforming to $\theta = z-\xi ({\bf R})$ yields:

\begin{eqnarray}
\lefteqn{t({\bf k}_{\rm in} \rightarrow {\bf k}_{\rm out}) \propto
\int dx \, dy \, d{\theta} \: V({\theta}) \, e^{i \,  \left[ 2 
\eta ({\bf R}) \, - \, \Delta {\bf K} \cdot {\bf R} \, - \, \Delta k^z \, (\xi
({\bf R}) \, + \, \theta) \right] } =} \nonumber \\
& & \underbrace{\int d{\bf R} \: e^{i \, [2 \eta ({\bf R}) \, - \, \Delta {\bf
K} \cdot {\bf R} \, - \, \Delta k^z \, \xi ({\bf R})]}}_{I_{\bf R}} \: \int
d{\theta} \: V({\theta}) e^{-i \, \Delta k^z \, \theta} .
\label{eq:1}
\end{eqnarray}

\noindent Clearly, only ${I_{\bf R}}$ can account for double
collisions. To proceed, it is convenient to perform a
stationary phase (SP) approximation \cite{Arfken}, with the purpose of
studying the singularities in the spectrum, i.e., the ``rainbow''
peaks. The SP yields:

\begin{equation}
\Delta {\bf K} = 2 \nabla \eta({\bf R}_0)\, - \, \Delta k^z \, \nabla \xi({\bf
R}_0) ,
\label{eq:xi-deltaK}
\end{equation}

\noindent and:

\[
{\left| {I_{\bf R}} (\Delta {\bf K})\right| }^2 \approx {(2 \pi)}^2 \,
{\left| 2 {{\partial^2 \, 
\eta({\bf R}_0)} \over {\partial \, {\bf R}^2}} \, - \, \Delta k^z
\,{{\partial^2 \, \xi({\bf R}_0)} \over {\partial \, {\bf
R}^2}}\right| }^{-1} .
\]

\noindent This singularity is a result of the crude SP approximation,
and is smoothed out in the full semiclassical expression
Eq.(\ref{eq:1}). However, it provides the condition from which the
inflexion point ${\bf R}_0$ is found:

\begin{equation}
{\left| 2 {{\partial^2 \, \eta({\bf R}_0)} \over {\partial \, {\bf
R}^2}} \, - \, 
\Delta k^z \, {{\partial^2 \, \xi({\bf R}_0)} \over {\partial \, {\bf
R}^2}}\right| } 
= 0 .
\label{eq:xi-R0}
\end{equation}

\noindent Eqs.(\ref{eq:xi-deltaK}),(\ref{eq:xi-R0}) together define
the {\em double-collision rainbow} (DCR) condition. After solving for
${\bf R}_0$ from Eq.(\ref{eq:xi-R0}), Eq.(\ref{eq:xi-deltaK})
determines the {\em position} of the DCR peak. It is similar to the SA
rainbow condition \cite{me:Heptamers}, $\Delta {\bf K} = 2 \nabla \eta({\bf R}_0)$. The
difference is thus in the second term, with the interesting feature
that it depends on the incidence momentum in the normal direction. A
direct prediction of the present analysis is therefore that {\em a DCR
peak is distinguished from a single-collision rainbow peak by a positional dependence on incidence
energy}. This effect should have a clear experimental signature. Further, the actual position of the peak depends on the
corrugation of the surface shape function at the inflexion
point. {\em This position thus provides valuable information on a surface
structural parameter.} A more complete analysis, including numerical
calculations of the double-collision rainbow intensities, is deferred
to a future publication.

\section{TDSCF Sudden Approximation for Atom-Surface Scattering}
\label{TDSCF-SA}

In this section I will present a generalized, full derivation of the SA. The
derivation is based on the original SA paper \cite{Benny:Sud1}, and its
extension to disordered surfaces \cite{Benny:disorder}, but instead of
considering the surface atoms as static, they are assumed to be
moving under the influence of a time-dependent self consistent field
(TDSCF). The result will be a TDSCF-SA, which is presented here for
the first time. Applications may include scattering from quantum vibrating
solid and liquid surfaces. A similar approach was pioneered by
Gerber \cite{Benny:unpublished}, but with the surfaces degrees of
freedom treated classically. A variation on this mixed
quantum-classical approach can be found in Ref. \cite{me:CAMP}.

\subsection{Preliminaries}
Consider a set of $N$ surface particles moving in a TDSCF, ${\bf
u}_n \!=\! (x_n,y_n)$ being the position of the $n^{\rm th}$ particle,
assumed constrained to the surface $z \!=\! 0$. Let $f_{\subkap_n}({\bf
u}_n,t) \!=\! \langle {\bf u}_n | \kap_n \rangle $ be the spatial
wavefunction for a surface particle with wavevector $\kap_n \!=\!
(\kappa^n_x,\kappa^n_y)$, solving the Schr\"{o}dinger equation

\[
H_n f_{\subkap_n}({\bf u}_n,t) = i\hbar {{\partial
f_{\subkap_n}({\bf u}_n,t)} \over {\partial t}} ,
\]

\noindent with the SCF Hamiltonian $H_n \!=\! v_n({\bf u}_n,t) \!+\!
T_n$, where 
$v_n$ is the SCF potential for the $n^{\rm th}$ particle and $T_n$ its
kinetic operator. A TDSCF separation based on the atoms as the
separable degrees of freedom may not be the best, as the residual
(their interaction) may become large due to collisions. Nevertheless,
for a liquid (the main application I have in mind) no better separable
coordinates are known. Moreover, it is possible to switch to lattice
modes (phonons) at a later point in the formalism (see below). Now,
from the completeness relation for $| {\bf u}_n \rangle $ and $| \kap_n
\rangle $,

\[
\int d{\bf u}_n \: | {\bf u}_n \rangle \langle {\bf u}_n | = 
\int d\kap_n \: | \kap_n \rangle \langle \kap_n | = 1 ,
\]

\noindent it follows that $f_{\kap_n}({\bf u}_n,t)$ is
orthogonal in both variables:

\ba
\int d{\bf u}_n \: f^*_{\subkap_n}({\bf u}_n,t) f_{\subkap'_n}({\bf
u}_n,t) = \langle \kap_n | \kap'_n \rangle = \delta(\kap_n-\kap'_n)
\label{eq:ortho1} \\
\int d\kap_n \: f^*_{\subkap_n}({\bf u}_n,t) f_{{\subkap}_n}({\bf 
u}'_n,t) = \langle {\bf u}_n | {\bf u}'_n \rangle = \delta({\bf
u}_n-{\bf u}'_n) .
\label{eq:ortho}
\ea

\noindent In addition the SCF separation of the surface particles
implies:

\[
\langle {\bf u}_n | {\bf u}_{n'} \rangle =
\langle \kap_n | \kap_{n'} \rangle = \delta_{n n'} .
\]

\subsection{Matrix-Diagonalization Sudden (MDS)}
\noindent Turning attention to the He particle, described by the
wavefunction $\psi$, the time-dependent Schr\"{o}dinger equation reads:

\be
[ T + V({\bf R},z;\{{\bf u}_n\}_N;t) ] \psi({\bf r}; \{{\bf u}_n\}_N; t) = 
i\hbar {{\partial \psi({\bf r}; \{{\bf u}_n\}_N; t)} \over {\partial
t}} ,
\label{eq:He}
\ee

\noindent where $T = (\hbar^2/2\mu)\nabla^2$ the kinetic energy
operator of the He particle, $V$ its potential energy which depends on $\{{\bf u}_n\}_N$,
the positions of the $N$ surface particles. The surface is assumed to be almost
frozen during the collision, so that the dependence of $V$ on $t$ is
very slow. This time-dependence arises from the slow motion of the surface
particles, whose dynamics I wish to include presently, in
order to develop a description which accounts for the details of the
energy transfer. It should be noted that it is inconceivable at this
point to imagine an experiment which is capable of resolving the
details of the energy transfer at the level of surface atoms in a liquid, which is
thermalized very quickly. However,
such experiments may become possible in the future (indeed, efforts
are underway in the context of atomic Ar clusters \cite{Buck:96}) so
that a formalism capable of describing them is desirable.

$V$ and the SCF potentials are not independent;
assuming that the influence of the He particle on the surface is small
and localized, it should be sufficient to evaluate $V$ at ${\bf r} \!=\!
{\bf 0}$, thus avoiding the need to find $\psi$ in order to calculate
the SCF potentials:

\be
v({\bf u}_n,t) =
\int \prod_{i \ne n} d{\bf u}_i \:
f^*_{\subkap_i}({\bf u}_i,t)
V({\bf 0};{\bf u}_i;t)
f_{\subkap_i}({\bf u}_i,t) .
\label{eq:v}
\ee

\noindent It is of course possible to average over $\psi$ as
well; then Eq.(\ref{eq:v}) should be regarded as the first iteration
in a TDSCF scheme which includes the He particle at later
iterations. I now assume that the SCF potentials have been obtained
and the problem is the solution of the Schr\"{o}dinger
equation (\ref{eq:He}). For this purpose it is convenient to expand
$\psi$ in a generalized time-dependent close-coupling form:

\be
\psi_{\bf k}({\bf r}; \{{\bf u}_n\}_N; t) =
\sum_{n=1}^N \int d\epsilon \int d{\bf q} \int d{\kap_n} \: e^{i({\bf
q}+{\bf K})\cdot {\bf R}} f_{\subkap_n}({\bf u}_n,t) \Phi_{\epsilon{\bf
q}\subkap_n}(z) e^{-i (E+\epsilon) t/\hbar}.
\label{eq:CC}
\ee

\noindent Here $\hbar {\bf k} = \hbar ({\bf K},k_z)$ is the momentum of the
incident He particle, $\hbar {\bf q} = \hbar (q_x,q_y)$ the
momentum transfer parallel to the surface, $E \!=\! (\hbar
k)^2/(2\mu)$ is the He incidence energy, and $\epsilon$ is the energy
transfer between the He and the surface. Since periodicity is not
assumed, the ${\bf q}$'s are unrestricted and continuous. Next the 
potential is taken to have the form:\footnote{$n,l$ are indices, not powers.}

\ba
V({\bf R},z;\{{\bf u}_n\}_N;t) =
\sum_{n=1}^N V^n({\bf R},z,{\bf u}_n,t) \\
\label{eq:V}
V^n({\bf R},z,{\bf u}_n,t) = \sum_{l=1}^L V^{n l}(z,t) Q^{n l}({\bf
R},{\bf u}_n,t) .
\label{eq:V^n}
\ea

\noindent No restrictions are placed at this point on the functions
$V^{n l}$ and $Q^{n l}$ or on their number $L$, so the expansion is
general (more so than in the other derivations of the SA,
Refs. \cite{Benny:Sud1,Benny:disorder}). It is however assumed, in
the spirit of SCF, that the He 
atom has only two-body interactions with the surface particles. The 
close-coupling equations will follow from inserting these expressions
for $\psi$ and $V$ into the Schr\"{o}dinger equation
(\ref{eq:He}). Since the surface particles are decoupled by virtue of
the SCF approximation, this results in $N$ equations of the form:

\ba
\int d\epsilon \int d{\kap_n} \:
\left[
{\hbar^2 \over 2\mu} \left( {\partial^2 \over {\partial z^2}} +
{\partial^2 \over {\partial {\bf R}^2}} \right) +
{\hbar^2 \over 2m_n} {\partial^2 \over {\partial {\bf u}_n^2}} -
\sum_{l=1}^L V^{n l}(z,t) Q^{n l}({\bf R},{\bf u}_n,t) + (E+\epsilon)
\right] &\times&
\nonumber \\
f_{\subkap_n}({\bf u}_n,t) e^{-i (E+\epsilon) t/\hbar} \int d{\bf q} \: e^{i({\bf q}+{\bf K})\cdot {\bf R}}
\Phi_{\epsilon{\bf q}\subkap_n}(z) &=& 0, \nonumber
\\
\label{eq:x1}
\ea

\noindent where the term $\partial f_{\subkap_n}({\bf u}_n,t)/\partial
t$ was neglected with respect to $(E\!+\!\epsilon)/\hbar \,
f_{\subkap_n}({\bf u}_n,t)$, in accordance with the assumption of slow
surface dynamics. \\
Multiplying throughout by $(\mu/\pi\hbar^2)\exp(-i{\bf
q}' \!\cdot\! {\bf R})$ and integrating over ${\bf R}$ yields, using $\int
d{\bf R} \: \exp(i {\bf q}\cdot {\bf R}) \!=\! 2\pi \delta({\bf q})$:

\ba
\int d\epsilon \left( \int d{\kap_n} \: \left[
{\partial^2 \over {\partial z^2}} - ({\bf q} +  {\bf K})^2 +
{\mu \over m_n} {\partial^2 \over {\partial {\bf u}_n^2}} +
k_\epsilon^2 \right]
f_{\subkap_n}({\bf u}_n,t) \Phi_{\epsilon{\bf q}\subkap_n}(z) - \right. 
\nonumber \\
\left. \int d{\kap_n} \:
\left[ \sum_{l=1}^L U^{n l}(z) Q^{n l}_{{\bf q},{\bf q}'}({\bf
u}_n) \right]
f_{\subkap_n}({\bf u}_n,t) \Phi_{\epsilon{\bf q}'\subkap_n}(z) \right) e^{-i
(E+\epsilon) t/\hbar} = 0 .
\label{eq:x2}
\ea

\noindent Here:
\ba
k_\epsilon^2 &\equiv& {2\mu \over \hbar^2} (E+\epsilon) \\
U^{n l}(z,t) &\equiv& {2\mu \over \hbar^2} V^{n l}(z,t) \\
Q^{n l}_{{\bf q},{\bf q}'}({\bf u}_n) &\equiv&
{1 \over 2\pi} \int d{\bf R} \: e^{i({\bf q}'-{\bf q})\cdot {\bf R}} Q_{n
l}({\bf R},{\bf u}_n) .
\label{eq:Q^nl}
\ea

\noindent Next the orthogonality of the surface particles'
wavefunctions is employed to eliminate the integrals over $\kap_n$ in
Eq.(\ref{eq:x2}). Multiplying throughout by $f^*_{{\subkap}'_n}({\bf
u}_n,t)\exp(i(E+\epsilon') t/\hbar)$ and integrating over ${\bf u}_n$
and $t$ yields, using Eq.(\ref{eq:ortho1}):

\ba
\hbar \left[
{\partial^2 \over {\partial z^2}} - ({\bf q} +  {\bf K})^2 + k_\epsilon^2
\right] \Phi_{\epsilon{\bf q}\subkap_n}(z) +
\int d\kap_n' \: F^n_{\epsilon\subkap_n,\epsilon'\subkap_n'}
\Phi_{\epsilon'{\bf q}\subkap_n'}(z) = \nonumber \\
\int d\kap_n' \: H^n_{\epsilon{\bf q}\subkap_n,\epsilon'{\bf
q}'{\subkap}_n'}(z) \Phi_{\epsilon'{\bf q}'\subkap_n'}(z) . 
\label{eq:x3}
\ea

\noindent Here:

\[
F^n_{\epsilon\subkap_n,\epsilon'\subkap_n'} \equiv {\mu \over m_n}
\int dt \: d{\bf 
u}_n \: f^*_{\subkap_n'}({\bf u}_n,t) {\partial^2 \over {\partial
{\bf u}_n^2}} f_{\subkap_n}({\bf u}_n,t)
e^{i(\epsilon'-\epsilon)t/\hbar}
\]

\noindent is (up to a constant) a component of the kinetic energy
matrix of the $n^{\rm th}$ surface particle associated with an $\epsilon
\rightarrow \epsilon'$ transition of the He particle, and:

\ba
H^n_{\epsilon{\bf q}\subkap_n,\epsilon'{\bf q}'\subkap_n'}(z) &\equiv& 
\sum_{l=1}^L U^{n l}(z,t) Q^{n l}_{\epsilon{\bf q}\subkap_n,\epsilon'{\bf
q}'{\subkap}_n'} \\
\label{eq:H}
Q^{n l}_{\epsilon{\bf q}\subkap_n,\epsilon'{\bf q}'\subkap_n'}
&\equiv& \int dt \:
d{\bf u}_n \: f^*_{\subkap_n'}({\bf u}_n,t) 
Q^{n l}_{{\bf q},{\bf q}'}({\bf u}_n) f_{\subkap_n}({\bf u}_n,t)
e^{i(\epsilon'-\epsilon)t/\hbar}
\label{eq:Q}
\ea

\noindent is a component of the matrix of the Fourier-transformed 
interaction between the He atom and the $n^{\rm th}$ surface
particle. Defining the diagonal ``energy operator'' ${\bf g}_\epsilon^2$ with
non-zero components

\be
g^2_{\epsilon\bf q} \equiv k_\epsilon^2 - ({\bf q} +  {\bf K})^2 ,
\label{eq:g^2}
\ee

\noindent Eq.(\ref{eq:x3}) can be rewritten as follows:

\ba
\int d\epsilon' \: d{\bf q}' \: d\kap_n' \: \left[
\delta[(\epsilon-\epsilon')/\hbar] \delta({\bf q}-{\bf q}')
\delta(\kap_n-\kap_n') \left(
{\partial^2 \over {\partial z^2}} + g^2_{\epsilon'{\bf q}'} \right)
\right. &+& \nonumber \\
\delta({\bf q}-{\bf q}') F^n_{\epsilon\subkap_n,\epsilon'\subkap_n'}
&-& \nonumber \\ 
\left. H^n_{\epsilon{\bf q}\subkap_n,\epsilon'{\bf q}'\subkap_n'}(z)
\right] \Phi_{\epsilon'{\bf q}'\subkap_n'}(z) &=& 0 .
\label{eq:x4}
\ea

\noindent This can be thought of as the $(\epsilon,{\bf q},\kap_n)$
component of a linear operator equation for the vector
$|\Phi^n\rangle$, whose components are the wavefunctions $\Phi_{\epsilon{\bf
q}\subkap_n}(z)$. Clearly then, the He kinetic energy operator on the first line of
Eq.(\ref{eq:x4}) is fully diagonal, the $n^{\rm th}$ surface particle
kinetic energy operator ${\cal F}^n$ on the second line is
diagonal in ${\bf q}$, and the interaction operator ${\cal H}^n$ on
the third line is not diagonal. In the Born approximation one
essentially retains only the fully diagonal (kinetic) term \cite{Taylor}. Ideally one would like to diagonalize the
complete operator 
acting on $|\Phi^n \rangle$, which would solve the scattering
problem. This is of course unrealistic, and here I will settle for
diagonalizing the {\em interaction} operator. It is to this end that
one introduces the Sudden approximation, which is usually written
as \cite{Benny:Sud1} :

\be
2\hbar k_z \gg \hbar |{\bf q}| .
\label{eq:sud}
\ee

\noindent This implies that the incident momentum in the
direction perpendicular to the surface ($\hbar k_z$) is large compared
with the momentum transfer for all the channels that significantly
contribute to scattering. Here it is necessary to add the assumption:

\be
E \gg \epsilon .
\label{eq:E-eps}
\ee

\noindent These conditions are satisfied for most present-day
experimental systems.
Using Eqs.(\ref{eq:sud}),(\ref{eq:E-eps}) one can neglect all but the
specular term in the energy operator, i.e., assume that:

\be
{\bf g}_\epsilon^2 \approx g^2_{0,\bf 0} {\cal I} = k_z^2 {\cal I} ,
\label{eq:sud1}
\ee

\noindent where ${\cal I}$ is the identity operator. Since
$g^2_{\epsilon{\bf q}} 
\!=\! k_z^2 \!-\! ({\bf q} \!+\! 2{\bf K}) \!\cdot\! {\bf q} \!+\!
2\mu\epsilon/\hbar^2$, it is seen 
that the real assumption made is somewhat stronger than
Eq.(\ref{eq:sud}):

\be
k_z^2 \gg |({\bf q} \!+\! 2{\bf K}) \!\cdot\! {\bf q}| .
\label{eq:sud2}
\ee

\noindent This implies that the incident beam should be nearly
perpendicular to the surface, a condition which is in fact sometimes
not met in practice, where the beam is often incident at 45$^\circ$
and the SA calculations have to be corrected accordingly
\cite{me:Ag-systems,me:SSL97,PRB:comment1}. The utility of the  
approximation Eq.(\ref{eq:sud1}) is that it will leave the energy
operator ${\bf g}^2$ diagonal after diagonalization of the interaction
operator in Eq.(\ref{eq:x4}). However, whereas this suffices
in the stationary surface case, here one is facing also the
operator ${\cal F}^n$, which is diagonal in ${\bf q}$ but not in
$\kap_n$ and $\epsilon$. For this reason it cannot be diagonalized
simultaneously 
with ${\cal H}^n$, since to diagonalize this operator it will be
necessary to rotate the ${\bf q}$ basis and hence introduce ${\bf
q}$-off-diagonal elements into ${\cal F}^n$. Nevertheless the problem
can be solved if one assumes as before when $\partial f_{\subkap_n}({\bf u}_n,t)/\partial
t$ was neglected, in the spirit of the SA, that {\em the
kinetic energy associated with the surface particles' motion is very
small compared to that of the incident He atom}. This is certainly
correct for solids, and also a reasonable assumption for liquids at
common experimental temperatures. Then it is possible to simply
neglect $F^n_{\epsilon\subkap_n,{\epsilon'\subkap_n'}}$ with respect
to $k_z^2$. With 
these approximations Eq.(\ref{eq:x4}) becomes, in operator form:

\ba
\left[
{\cal I} \left( {\partial^2 \over {\partial z^2}} +
k_z^2 \right) - {\cal H}^n
\right] |\Phi^n \rangle = 0 .
\label{eq:x5}
\ea

\noindent This represents a set of coupled equations for scattering of
the He atom by the $n^{\rm th}$ surface particle into the available
channels indexed by $\epsilon$, ${\bf q}$ and $\kap_n$. To uncouple the
set one needs a unitary matrix ${\cal B}^n$ that diagonalizes {\em all}
the ($z$-independent) ${\cal Q}^{n l}$ matrices [whose components are
defined in 
Eq.(\ref{eq:Q})]. That a single ${\cal B}^n$ can diagonalize all $L$
${\cal Q}^{n l}$ matrices is not {\it a priori} obvious. I will assume 
for the moment that this is true and come back to it later [see
Eq.(\ref{eq:QR})]. Then:

\be
{\cal Q}_D^{n l} = {\cal B}^n {\cal Q}^{n l} ({\cal B}^n)^{-1} ,
\label{eq:Q^D}
\ee

\noindent with matrix elements

\[
\langle \epsilon'{\bf q}'\kap_n' | {\cal Q}_D^{n l} | \epsilon{\bf
q}{\kap}_n \rangle =
\langle \epsilon{\bf q}\kap_n | {\cal Q}_D^{n l} | \epsilon{\bf
q}{\kap}_n \rangle \delta(\epsilon-\epsilon') \delta({\bf q}-{\bf q}')
\delta(\kap_n-\kap_n') .
\]

\noindent Since ${\cal B}^n$ leaves the energy matrix diagonal in view
of Eq.(\ref{eq:sud1}), and commutes with purely $z$-dependent
operators, after application of ${\cal B}^n$ Eq.(\ref{eq:x5}) becomes:

\ba
\left[
{\cal I} \left( {\partial^2 \over {\partial z^2}} +
k_z^2 \right) - \sum_{l=1}^L U^{n l}(z,t) {\cal Q}_D^{n l}
\right] |\tilde{\Phi}^n \rangle = 0 ,
\label{eq:x6}
\ea

\noindent where

\[
|\tilde{\Phi}^n \rangle = {\cal B}^n |\Phi^n \rangle .
\]

\noindent The system of equations (\ref{eq:x6}) is  {\em uncoupled}
since the operator acting on $|\tilde{\Phi}^n \rangle$ is now
diagonal. Physically, this corresponds to the fact that there are no
transitions between the decoupled channels labeled by $\epsilon$,
${\bf q}$ and 
$\kap_n$ in the new system. Thus the scattering matrix
${\cal S}_D$ that arises from Eq.(\ref{eq:x6}) is diagonal and
represents elastic scattering in each of the decoupled channels. Its
elements therefore satisfy:

\be
\langle {\epsilon'\bf q}'\kap_n' | {\cal S}_D^n | \epsilon{\bf
q}{\kap}_n \rangle(t) = 
e^{2i \eta^n_{\epsilon{\bf q}\subkap_n}(t)} \delta(\epsilon-\epsilon') \delta({\bf q}-{\bf q}')
\delta(\kap_n-\kap_n') ,
\label{eq:S^D}
\ee

\noindent where $\eta^n_{\epsilon{\bf q}\subkap_n}(t)$ is the elastic
scattering phase shift determined for the ${\epsilon{\bf q}\kap_n}$
channel from Eq.(\ref{eq:x6}). The time-dependence of the S-matrix
element may appear strange (after all the S-matrix describes the entire
scattering process from $t \!=\! -\infty \rightarrow \infty$), but it
should be remembered that ${\cal S}_D^n$ is not the physical S-matrix;
it still needs to be transformed back to the physical frame, which
will remove the time-dependence (see below). $\eta^n_{\epsilon{\bf
q}\subkap_n}(t)$ can be evaluated explicitly in the WKB approximation:

\[
\eta^n_{\epsilon{\bf q}\subkap_n}(t) = 
\int_{\xi({\epsilon{\bf q}\subkap_n};t)}^{\infty} dz \: \left[ \left(
k_z^2 - \sum_{l=1}^L U^{n l}(z,t) \langle \epsilon{\bf q}\kap_n | {\cal
 Q}_D^{n l} | \epsilon{\bf q}\kap_n \rangle \right)^{1/2} -k_z^2
\right]
+ k_z \xi({\epsilon{\bf q}\subkap_n};t) ,
\]

\noindent where $\xi({\epsilon{\bf q}\subkap_n};t)$ is the classical turning
point associated with the ${\epsilon{\bf q}\kap_n}$ decoupled channel,
i.e., is the $z$ value that solves

\[
\sum_{l=1}^L U^{n l}(z,t) \langle \epsilon{\bf q}\kap_n | {\cal
Q}_D^{n l} | \epsilon{\bf q}\kap_n \rangle = k_z^2 .
\]

\noindent To complete the solution of the scattering problem within
this {\em Matrix Diagonalization Sudden} (MDS) approximation, as
mentioned above one needs
to transform ${\cal S}_D^n$ back to the physical scattering matrix
${\cal S}^n$. It can be shown that this is accomplished by the inverse
unitary transformation \cite{Goodman:70}:

\be
{\cal S}^n = ({\cal B}^n)^{-1} {\cal S}_D^n {\cal B}^n .
\label{eq:S}
\ee

\noindent That this transformation removes the time-dependence will be
demonstrated explicitly in the coordinate representation Sudden,
below. Equations (\ref{eq:S^D})-(\ref{eq:S}) together provide the
amplitude for scattering by the $n^{\rm th}$ surface particle. In the
present approach the contributions from all these particles are
considered to be additive, so that the full transition amplitude from
channel $\epsilon{\bf q}$ to $\epsilon'{\bf q}'$ is:

\be
\langle \epsilon'{\bf q}' | {\cal S} | \epsilon{\bf q} \rangle =
\sum_{n=1}^N \langle \epsilon'{\bf q}'\kap_n' | {\cal S}^n | \epsilon{\bf
q}\kap_n \rangle .
\label{eq:Sfinal}
\ee

\noindent Note that these are generally complex-valued amplitudes, so
that there may be interference between scattering from different
surface particles. The representation of the S-matrix as a sum over
particles may be questioned. However, since He at typical scattering
energies has a wavelength between 0.5\AA and 5\AA whereas in a
liquid the average inter-particle separation is at least on the order
of the corresponding solid's lattice constant, i.e., several \AA, only at
extremely low incidence energies this approximation is expected to
break down. This however, is anyway beyond the range of validity of
the SA. Moreover, if a representation in terms of lattice modes is
required, it can be transformed to by representing the particle
displacements in terms of creation ($a_i^+$) and annihilation ($a_i$)
operators \cite{Benny:DW}:

\ba
{\bf u}_n = \sum_{i=1}^N \left({{\mbox{\boldmath $\alpha$}}}_{n i} a_i^+ + {{\mbox{\boldmath $\alpha$}}}_{n
i}^* a_i \right) \nonumber \\
{{\mbox{\boldmath $\alpha$}}}_{n i} = (2m_n N \omega_i)^{-1/2} {\bf e}_i e^{i {\bf K}_i
\cdot {\bf R}_n^{(0)}} \nonumber ,
\ea

\noindent where ${\bf R}_n^{(0)}$ denotes the equilibrium position of
the $n^{\rm th}$ surface atom and $\omega_i$, ${\bf K}_i$, and ${\bf
e}_i$ denote the frequency, wave vector, and polarization vector of
mode $i$, respectively \cite{Ghatak:book72}.

\subsection{Coordinate Representation Sudden (CRS)}
In practice the MDS has almost not been used; instead one can
transform to the coordinate representation to obtain the more
convenient and popular {\em 
Coordinate Representation Sudden} (CRS). To accomplish this it should
be noted first that {\em if they are not truncated},
the ${\cal Q}^{n l}$ are trivially diagonal in the coordinate
representation, since they are made of matrix elements of the
functions $Q^{n l}({\bf R},{\bf u}_n,t)$ \cite{PRB:comment2}.
Therefore, in this representation:

\be
\langle t{\bf R} {\bf u}_n | {\cal Q}^D_n |t' {\bf R}' {\bf u}_n' \rangle =
Q^{n l}({\bf R},{\bf u}_n,t) \delta(t'-t) \delta({\bf R}-{\bf R}')\delta({\bf
u}_n-{\bf u}_n') .
\label{eq:QR}
\ee

\noindent At this point it should be clear when a single diagonalizing
matrix ${\cal B}^n$, required for the MDS, can be found for all ${\cal
Q}^{n l}$ matrices: since they are all simultaneously diagonal in the
coordinate representation if they are not truncated, they all commute,
and are therefore simultaneously diagonalizable in any other basis as
well. The same reasoning can be applied to the scattering matrix,
which was found to be diagonal in the same basis as the ${\cal Q}^{n
l}$ [Eqs.(\ref{eq:Q^D}), (\ref{eq:S^D})]. Thus ${\cal S}^n$ is also
diagonalized by transforming to the coordinate representation:

\be
\langle t{\bf R} {\bf u}_n | {\cal S}_D^n | t'{\bf R}' {\bf u}_n' \rangle =
S^n({\bf R},{\bf u}_n) \delta(t'-t) \delta({\bf R}-{\bf R}')\delta({\bf
u}_n-{\bf u}_n') .
\label{eq:SR}
\ee

\noindent Here $S^n({\bf R},{\bf u}_n)$ is found from the solution
of the coordinate representation version of Eq.(\ref{eq:x6}) (obtained
by inserting the unit operator $\int dt'\: d{\bf R}' \: d{\bf u}_n' \: |
t' {\bf R}' {\bf u}_n' \rangle \langle t' {\bf R}' {\bf u}_n' |$ before
$|\tilde{\Phi}' \rangle$ and
multiplying on the left by $\langle t {\bf R} {\bf u}_n |$):

\be
\left[
\left( {\partial^2 \over {\partial z^2}} +
k_z^2 \right) - V^n({\bf R},z,{\bf u}_n,t) \right] \tilde{\Phi}_{t{\bf R} {\bf
u}_n}(z) = 0 ,
\label{eq:x7}
\ee

\noindent where $V^n({\bf R},z,{\bf u}_n,t)$ was defined in
Eq.(\ref{eq:V^n}), and $\tilde{\Phi}_{t{\bf R} {\bf u}_n}(z) \!=\!
\langle {t\bf R} {\bf u}_n | {\cal B}^n | \Phi^n \rangle $ depends
parametrically on $t$, ${\bf
R}$ and ${\bf u}_n$. Eq.(\ref{eq:x7}) shows that the SA treats the
coordinates $t$, ${\bf R}$ and ${\bf u}_n$ in an {\em adiabatic}
approximation. One can now write:

\[
\langle \epsilon'{\bf R}' {\bf u}'_n | S_D^n | \epsilon{\bf R} {\bf
u}_n \rangle = 
e^{2i \eta^n_\epsilon({\bf R},{\bf u}_n,t)} \delta(\epsilon-\epsilon')
\delta({\bf R}'-{\bf R}) \delta({\bf u}'_n-{\bf u}_n) ,
\]

\noindent with the elastic-scattering phase shift evaluated from
Eq.(\ref{eq:x7}) in the WKB approximation as:

\[
\eta^n_\epsilon({\bf R},{\bf u}_n,t) = 
\int_{\xi({{\bf R},{\bf u}_n},t)}^{\infty} dz \: \left[ \left(
k_z^2 - V^n({\bf R},z,{\bf u}_n,t) \right)^{1/2} -k_z^2 \right]
+ k_z \xi({{\bf R},{\bf u}_n},t) .
\]

\noindent Here $\xi({{\bf R},{\bf u}_n},t)$ is the classical turning point
pertaining to Eq.(\ref{eq:x7}). Next, the coordinate
representation S-matrix has to be transformed back to the physical
one, again using Eq.(\ref{eq:S}):

\ba
\lefteqn{
\langle \epsilon'{\bf q}'\kap_n' | {\cal S}^n | \epsilon{\bf q}\kap_n
\rangle =} \nonumber \\
&&
\int dt'\: d{\bf R}'\: d{\bf u}_n'\: dt\: d{\bf R}\: d{\bf u}_n\:
\langle \epsilon'{\bf q}'\kap_n' | ({\cal B}^n)^{-1} | t{\bf R} {\bf u}_n \rangle
\langle t{\bf R} {\bf u}_n | {\cal S}_D^n | t'{\bf R}' {\bf u}_n' \rangle
\langle t'{\bf R}' {\bf u}_n' | {\cal B}^n | \epsilon{\bf q}\kap_n
\rangle .
\label{eq:S1}
\ea

\noindent The integrals over the primed variables are eliminated with
the help of Eq.(\ref{eq:SR}); in addition one needs the ${\cal B}^n$ matrix
elements. On the one hand:

\ba
\lefteqn{
Q^{n l}_{\epsilon{\bf q}\subkap_n,\epsilon'{\bf q}'\subkap_n'} =
\langle \epsilon'{\bf q}'\kap_n' | {\cal Q}^{n l} | \epsilon{\bf
q}{\kap}_n \rangle = } \nonumber \\
&&
\langle \epsilon'{\bf q}'\kap_n' | ({\cal B}^n)^{-1} {\cal Q}_D^{n l}
{\cal B}^n | \epsilon{\bf q}{\kap}_n \rangle = \nonumber \\
&&
\int dt' \: d{\bf R}'\: d{\bf u}_n'\: dt \: d{\bf R}\: d{\bf u}_n\:
\langle \epsilon'{\bf q}'\kap_n' | ({\cal B}^n)^{-1} | t {\bf R} {\bf
u}_n \rangle \langle t {\bf R} {\bf u}_n | {\cal Q}^D_n | t' {\bf R}' {\bf
u}_n' \rangle 
\langle t' {\bf R}' {\bf u}_n' | {\cal B}^n | \epsilon{\bf q}\kap_n \rangle =
\nonumber \\
&&
\int dt \: d{\bf R}\: d{\bf u}_n\:
\left\{ \langle t {\bf R} {\bf u}_n | {\cal B}^n | \epsilon'{\bf q}' \kap_n' \rangle
\right\}^*
Q^{n l}({\bf R},{\bf u}_n,t) 
\left\{ \langle t {\bf R} {\bf u}_n | {\cal B}^n | \epsilon{\bf
q}\kap_n \rangle \right\} , \nonumber
\ea

\noindent where the last line follows from Eq.(\ref{eq:QR}) and the
unitarity of ${\cal B}^n$. On the other hand one has from
Eqs.(\ref{eq:Q^nl}),(\ref{eq:Q}):

\ba
\lefteqn{
Q^{n l}_{\epsilon{\bf q}\subkap_n,\epsilon'{\bf q}'\subkap_n'} = }
\nonumber \\
&&
{1 \over 2\pi} \int dt \: d{\bf R}\: d{\bf u}_n\:
\left\{ f_{\subkap_n',t}({\bf u}_n) e^{-i( \epsilon't/\hbar - {\bf
q}'\cdot {\bf R} )} \right\}^*
Q^{n l}({\bf R},{\bf u}_n,t) 
\left\{ f_{\subkap_n,t}({\bf u}_n,t) e^{-i (\epsilon t/\hbar - {\bf
q}\cdot {\bf R} )} \right\} , \nonumber
\ea

\noindent which shows that ${\cal B}^n$ is nothing but the matrix of
transformation between the momenta and coordinates:

\[
\langle t {\bf R} {\bf u}_n | {\cal B}^n | \epsilon {\bf q}\kap_n \rangle =
{1 \over 2\pi} f_{\subkap_n}({\bf u}_n,t) e^{i{\bf q}\cdot {\bf R}}
e^{-i\epsilon t/\hbar} .
\]

\noindent Inserting this into Eq.(\ref{eq:S1}) one obtains finally:

\ba
\lefteqn{
\langle \epsilon' {\bf q}'\kap_n' | {\cal S}_n | j {\bf q}\kap_n
\rangle =} \nonumber \\
&&
{\alpha \over 4\pi^2} \int dt \: d{\bf R}\: d{\bf u}_n\:
\left\{ f^*_{\subkap_n'}({\bf u}_n, t)
e^{-i({\bf q}' \cdot {\bf R} -
\epsilon' t/\hbar)} \right\}
e^{2i \eta_\epsilon^n({\bf R},{\bf u}_n,t)}
\left\{ f_{\subkap_n}({\bf u}_n, t)
e^{i({\bf q} \cdot {\bf R} - \epsilon t/\hbar)} \right\} ,
\label{eq:x8}
\ea

\noindent for the amplitude to scatter off the $n^{\rm th}$ surface
particle, accompanied by a momentum transfer $\hbar{\bf q} \rightarrow
\hbar{\bf 
q}'$ and energy transfer $\epsilon_{j} \rightarrow
\epsilon'$ by the He atom, and a change $\hbar{\kap}_n
\rightarrow \hbar{\kap}_n'$ in the momentum of the surface particle. As in the
MDS, the full amplitude for the scattering of the He atom is given by
the sum over all surface particles, Eq.(\ref{eq:Sfinal}). The
normalization factor $\alpha$ is found from the conservation of the
total scattering probability:

\be
\sum_n \int d\epsilon' \: d{\bf q}' \: d{\kap}_n' |\langle \epsilon' {\bf
q}'\kap_n' | {\cal S}_n | j {\bf q}\kap_n \rangle|^2 = 1 .
\label{eq:norm}
\ee

\noindent This completes the derivation of the TDSCF-SA, which is
given in final form by Eqs.(\ref{eq:Sfinal}), (\ref{eq:x8}) and
(\ref{eq:norm}). {\em The novel 
feature in 
the present result is the appearance of the time dependence and the
explicit accounting for the change undergone by the surface particles};
for comparison see again the usual SA expression 
Eq.(\ref{eq:Sud}). 

The time dependence in Eq.(\ref{eq:x8}) can be made even more explicit by
expanding the wave function of the surface particle in terms of its
energy eigenstates: 

\[
|\kap_n \rangle = \sum_{j_{{}_n}} c_{j_{{}_n}} | j_n \rangle e^{-i
E_{j_{{}_n}}t/\hbar} ,
\]

\noindent or, by multiplying on the left by $\langle {\bf u}_n |$, in
the coordinate representation:

\[
f_{\subkap_n}({\bf u}_n; t) = \sum_{j_{{}_n}} c_{j_{{}_n}}
\phi_{j_{{}_n}}({\bf u}_n) e^{-i E_{j_{{}_n}}t/\hbar} ,
\]

\noindent where $\phi_{j_{{}_n}}({\bf u}_n) = \langle {\bf u}_n | j_n
\rangle$. Plugging this into Eq.(\ref{eq:x8}) one obtains the matrix
element:

\ba
\langle \kap_n' | e^{2i \eta^n_\epsilon({\bf R},{\bf u}_n,t)} | \kap_n \rangle &\equiv&
\int d{\bf u}_n\:
f^*_{\subkap_n'}({\bf u}_n; t) e^{2i \eta^n_\epsilon({\bf R},{\bf u}_n,t)}
f_{\subkap_n}({\bf u}_n; t) \nonumber \\
&=& \sum_{j_{{}_n},j_n'} c'^*_{j_n'} c_{{}_{j_{{}_n}}} e^{i
(E_{j_n'}-E_{j_{{}_n}})t/\hbar} M_{{j_n'}{j_{{}_n}}}({\bf R},t) ,
\label{eq:x9}
\ea

\noindent where:

\[
M_{{j_n'}{j_{{}_n}}}({\bf R},t) = \int d{\bf u}_n\:
\phi^*_{j_n'}({\bf u}_n) e^{2i \eta^n_\epsilon({\bf R},{\bf u}_n,t)}
\phi_{j_{{}_n}}({\bf u}_n) .
\]

\noindent The $M_{{j_n'}{j_{{}_n}}}$ are the ${\bf R}$-dependent amplitudes
for a transition from energy eigenstate $| j_n \rangle$ to $| j_n'
\rangle$ for the $n^{\rm th}$ surface particle in the ``$t^{\rm th}$''
configuration. The off-diagonal
elements, with $ j_n' \ne j_n $, correspond to inelastic transitions
due to the collision with the He atom, as can be seen by the presence of the
oscillatory term with the $E_{j_n'}-E_{j_{{}_n}}$ factor in
Eq.(\ref{eq:x9}). {\em This demonstrates explicitly how the TDSCF-SA
accounts for energy transfer from/to the liquid surface}. The $M_{{j_n'}{j_{{}_n}}}$ have been calculated
analytically (under 
certain simplifying assumptions) for the harmonic oscillator case by
Schinke and Gerber \cite{Benny:DW}, where they give rise to the
Debye-Waller factor. The present formulation is more
general as it is not restricted to phonons, and explicitly includes
the energy transfer to/from the He atom. This novel feature should
make the present result useful in a comparison with inelastic
He scattering data from liquids, to be undertaken in a future publication.

\section{Summary and Conclusions}
\label{conclusions}

This paper has introduced two novel extensions of the Sudden
Approximation (SA) for atom scattering from surfaces, namely: (1)
scattering in the presence of double collisions between the impinging
atom and the surface (the Iterated Sudden: IS), and (2) scattering
from a time-dependent quantum liquid surface (the time-dependent self
consistent field SA: TDSCF-SA).

The IS is a natural generalization of the SA, which takes
double-collision events between the atom and the surface into
account. An exact version was presented, suitable primarily for
numerical applications, along with a more approximate version, which
can be used for analytical work. The exact version is only slightly
more computationally expensive than the original SA. It was shown that the IS
predicts the presence of new rainbow peaks in the angular intensity
distribution, which arise due to double collisions, and are absent in
the SA. These rainbow peaks can be experimentally distinguished from
those arising from 
single-collisions by their dependence on incidence energy. The IS
seems to properly generalize the SA, which was the 
goal in its derivation. It is of interest to apply the IS to a
realistic problem, e.g. a CO adsorbate on a Pt(111) surface, and
compare the IS predictions to those of the SA and exact
calculations.

The TDSCF-SA generalizes the SA by taking into account the dynamics of
the surface particles. The latter are treated as independent, with
each particle moving in an average potential determined by all the
others, and the incident atom. The resulting expression for the
scattering amplitude depends on the energy transfer between the
incident atom and the surface, and it is this inelastic contribution
which constitutes the main generalization of the original SA
approximation. The TDSCF-SA could find applications in simulations
related to the novel experiments on inert atom scattering from molten
metal liquid surfaces, such as those by Nathanson and
co-workers \cite{King:93,Ronk:96,Nathanson:96}.

Theoretical applications of the methods developed here will be
undertaken in the future, but it is 
hoped above all that this work will stimulate experimentalists to
further utilize inert atom scattering in the study of increasingly
complex surface disorder. The results presented here suggest that such
experiments can reveal a wealth of information concerning disordered
surface structure, be it in the atomic-level details of isolated
adsorbates, the dynamics and time-dependent structure of quantum
liquid surfaces, or the statistics of randomly corrugated surfaces.

\section*{Acknowledgements}

This work was carried out while the author was with the Physics
Department and the Fritz Haber Center for Molecular Dynamics at the
Hebrew University of Jerusalem, Givat Ram, Jerusalem 91904,
Israel. Numerous helpful discussions with Prof. R. Benny Gerber,
without whom this work could not have been completed, are gratefully
acknowledged.

\appendix

\section{Connection of the Iterated Sudden Expressions with the
Single-Collision Case}

\subsection{Wave-Function Iterated Sudden}
\label{app1}

It is interesting to rewrite Eq.(\ref{eq:WF-t}) in a form which makes the
connection with the single-collision case, i.e., the ordinary SA, more
transparent. Equation (\ref{eq:general-t}) can be rewritten as:

\begin{eqnarray}
\lefteqn{t^j ({\bf k}_{\rm in} \rightarrow {\bf k}_{\rm out}) =} \nonumber \\
& & {{\alpha} \over {{(2 \pi)}^{3 \over 2}}} \: \int d{\bf q} \: \int d{\bf
r} \: e^{i \,  \left[ ({\bf K}_{\rm in}-{\bf q}) \cdot {\bf R} -
k_{\rm out}^z \, z 
\right] } \, \Phi_{\bf R}^j (z) \int d{\bf R}' \: V({\bf R}',z) e^{i
\,  \left[ ({\bf q}-{\bf K}_{\rm out}) \cdot {\bf R}' \right] } .
\label{eq:DC}
\end{eqnarray}

\noindent The first term in the last expression is, up to a constant
factor, exactly the 
transition amplitude from a Sudden state $| \psi_{\rm in}^j \rangle$ to an {\em
intermediate} state $\langle {\bf q}, k_{\rm out}^z |$, as is shown next by
calculating the $\langle {\bf q}, k_{\rm out}^z |$ component of the SA
wave-function:

\begin{eqnarray}
\lefteqn{\langle {\bf q}, k_{\rm out}^z | \psi_{\rm in}^j \rangle =
\int d{\bf r} \: 
\langle {\bf q}, k_{\rm out}^z | {\bf r} \rangle \, \langle {\bf r} |
\psi_{\rm in}^j 
\rangle =} \nonumber \\
& & \int d{\bf q}' \; \int d{\bf r} \: \Phi_{{\bf q}'}^j (z) \, e^{i\, ({\bf
K}_{\rm in} + {\bf q}') \cdot {\bf R}} \, e^{-i({\bf q},k_{\rm out}^z)
\cdot {\bf r}} 
\, {(2 \pi)}^{-{3 \over 2}} = \nonumber \\
& & \sqrt{2 \pi} \, \alpha \: \int d{\bf r}\: \Phi_{\bf R}^j (z) \, e^{i\,
\left[ ({\bf K}_{\rm in}-{\bf q}) \cdot {\bf R} \, - \, k_{\rm out}^z
\, z \right] } , \nonumber
\end{eqnarray}

\noindent which is indeed the term mentioned above. Having verified
this, the double-collision expression obtained within the IS,
Eq.(\ref{eq:DC}), can be interpreted as the transition amplitude from
an initial state $| \psi_{\rm in}^j
\rangle$ (after a first collision) into an intermediate state $\langle {\bf
q}, k_{\rm out}^z |$, multiplied by a Born-type propagator $P \equiv
\int d{\bf R}' \: V({\bf R}',z) \, \exp\left[ i \, (({\bf q}-{\bf
K}_{\rm out}) \cdot {\bf R}') \right] $, which takes the particle to
the final state $\langle {\bf k}_{\rm out} |$ (after a second
collision), integrated over all intermediate diffraction vectors ${\bf
q}$. The ``propagator'' $P$ depends on $z$ through $V({\bf
R}',z)$. Thus the first and second collisions are coupled not only by
the intermediate momentum vector, but also by the (non-adiabatic)
$z$-coordinate.  As shall be shown shortly, this coupling is closely
connected to the distinction between off-shell and on-shell
amplitudes, and can be used to determine when an off-shell amplitude
can be approximated by an on-shell one.

\subsection{Phase-Shift Iterated Sudden}
\label{app2}

Consider Eq.(\ref{eq:on-shell}): the expression in parentheses is, according to
Eq.(\ref{eq:Sud}), just $\langle {\bf q} | S^{\rm sud} | {\bf K}_{\rm
in}
\rangle$, i.e., the single collision 
amplitude for the transition from $| {\bf K}_{\rm in} \rangle$ to the
intermediate state $\langle {\bf q} |$, in the SA. Thus
Eq.(\ref{eq:on-shell}) can be interpreted as expressing the transition
amplitude from an initial state $| {\bf K}_{\rm in} \rangle$ to a
final state $| {\bf k}_{\rm out} \rangle$, after passing through an
intermediate state $| {\bf q} \rangle$. The expression in parentheses
in Eq.(\ref{eq:on-shell}) then expresses the first collision, which is
subsequently multiplied by a Born type propagator, expressing the
second collision: the transition from $\langle {\bf q} |$ to $| {\bf
k}_{\rm out}
\rangle$. But as opposed to the situation in the Wave-Function expression
Eq.(\ref{eq:DC}), the two collision events in Eq.(\ref{eq:on-shell})
are {\em decoupled} in the $z$-direction. This is due to the on-shell
approximation used in deriving in Eq.(\ref{eq:on-shell}). Hence one
would expect an on-shell approximation to be valid as long as the
behavior of the particle in the $z$-direction is primarily determined by
the initial $z$-component of the particle's wave-vector. For in this
case, the influence of the potential on the propagation in the
$z$-direction is relatively negligible. This is in the spirit of the
high-energy SA.

\begin{thebibliography}{10}
\bibitem{PRB:comment2}
This non-truncation is a somewhat ambiguous requirement in the context of the
  SA: on the one hand it is in the spirit of the high-energy sudden since it
  corresponds to the existence of (infinitely) many open channels. On
the other hand Eq.(\ref{eq:sud}) requires some channels to be closed.
\end{thebibliography}
\end{document}